# Location Estimation with Reactive Routing in Resource Constrained Sensor Networks

S. Sarangi[1] and S. Kar[2]

[1]Bharti School of Telecommunication Technology
and Management, IIT Delhi,
Hauz Khas, New Delhi 110016 INDIA
sanat.sarangi@gmail.com

[2]Bharti School of Telecommunication Technology
and Management, IIT Delhi,
Hauz Khas, New Delhi 110016 INDIA
subrat@ee.iitd.ac.in

**Abstract**

Routing algorithms for wireless sensor networks can be broadly divided into two classes - proactive and reactive. Proactive routing is suitable for a network with a fixed topology. On the other hand, reactive routing is more suitable for a set of mobile nodes where routes are created on demand and there is not much time to evaluate the worthiness of a route, the prime concern being reachability due to constantly changing node positions. Sensor networks route events of interest from source(s) to destination(s) where appropriate actions could be taken. However, with mobile sensor nodes, it is not only important to know the events but the location of the nodes generating the events. Most sensor nodes are not equipped with expensive GPS or accurate RSSI computation hardware to aid localization. Keeping these in view, we propose a modified reactive routing algorithm, with added support for localization, to localize mobile sensor nodes on the basis of information received from fixed sensor nodes during mutual exchange of routing control packets. The accuracy of localization depends on the ratio of the number of fixed nodes to the number of mobile nodes and the topology of the fixed nodes. A typical application scenario would be a mix of mobile nodes and fixed nodes, where fixed nodes know their absolute location and the location of mobile nodes is derived from the fixed nodes, in step with the reactive routing protocol in action. The modified algorithm would be suitable for deployments where the approximate position of a mobile node (i.e. the event location) is required but there is no external support infrastructure available for localization.

## 1 INTRODUCTION

In this paper, we propose an extension, LORECOS, to a popular reactive routing protocol Ad-Hoc On Demand Distance Vector Routing (AODV) [1] to perform joint localization and routing of events in a wireless sensor network consisting of both fixed and mobile nodes. [2] summarizes a large number of on-demand routing protocols. We choose AODV as our routing algorithm since it is widely used due to its architectural simplicity and low routing overhead unlike DSR [3] which requires each packet to have full routing information instead of just the next hop to which the packet has to be forwarded. Moreover, a robust RFC [4] for AODV is also available which acts as a ready reference for carrying out implementation and enhancements. In [5] it has been shown that AODVjr, a considerably simpler and trimmed down version of AODV is able to give results comparable to AODV. We use the details in [4] and the guidelines in [5] to simulate AODVjr and add to it, our extension LORECOS to create an algorithm for joint localization and routing. We also discuss how the principles of LORECOS could be used with AODV.

In [6], coarse grained localization with a fixed-grid setup has been discussed, but, deploying a fixed grid of nodes may neither always be feasible nor desirable when done only for the purpose of localizing a set of mobile nodes due to the cost incurred on the additional nodes to form the grid itself. Coarse-grained localization with periodic beaconing by nodes on the fixed grid is a colossal wastage of communication power if the only information needed is the location of a set of sensor nodes moving in the grid. Instead, energy for localization could be minimized if localization is done opportunistically (on-demand) when an event is detected and routed through the network to the destination. Further, one of the most common sensor network deployments consists of a set of sensor nodes sending events over the network to a single destination (gateway) connected to a decision support system (DSS) for necessary action. Localization has broadly two key objectives - self-localization (i.e. the node being aware of its own location) and communication of the estimated location to the gateway. We present our results keeping these factors in view so that the advantages of deploying LORECOS on resource constrained nodes can be well appreciated.

Location Estimation with Reactive Routing in Resource Constrained Sensor Networks

## 2 NOTATION

We use the notation and abbreviations given in Table 1 for our discussion:

*Table 1   Notation*

| Abbreviation | Expanded Form | Abbreviation | Expanded Form | Abbreviation | Expanded Form |
| --- | --- | --- | --- | --- | --- |
| RREQ | AODV Route Request packet | MAE | Mean absolute error(m) in localization | HRQ | Hello Request packets |
| RREP | AODV Route Reply packet | RMSE | Root mean squared error(m) in localization | HRP | Hello Reply packets |
| RQU | RREQ packet without location | UN | A fixed (white) node not knowing its location | CLOC | CENTROID-LOC, LORECOS |
| RQK | RREQ packet with location | KN | A fixed (black) node knowing its location | HLOC | HELLO-LOC, LORECOS |
| RPU | RREP packet without location | UN-M | A mobile (white) node not knowing its location | DLOC | DISCOVERED-LOC, LORECOS |
| RPK | RREP packet with location | K(or U)N-x | K(or U)N with id=x | EVL | Event packet with location |

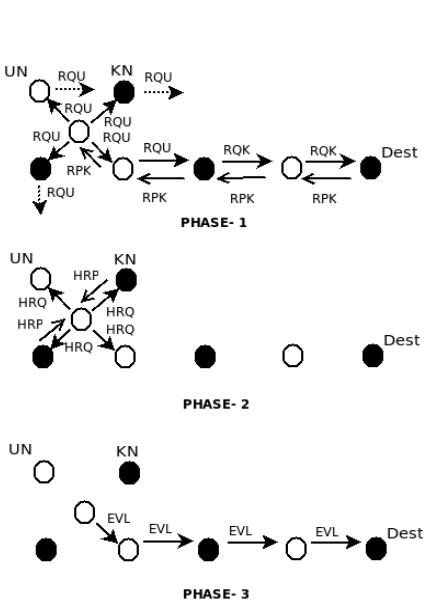

*Figure 1   Phases of LORECOS-AODVjr*

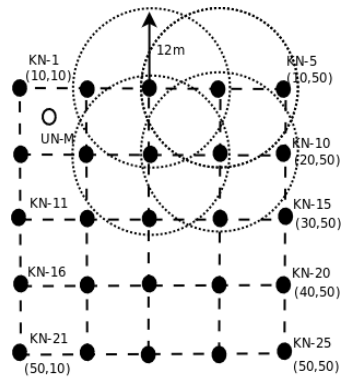

*Figure 2   Topology 1*

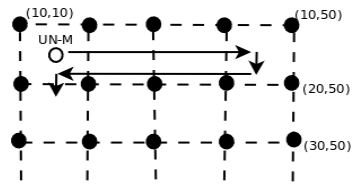

*Figure 3   Mobility Model*

## 3 LORECOS

Before going to LORECOS, it is imperative that we discuss AODV and AODVjr from which LORECOS is derived. AODVjr, a scaled down version of AODV does not use sequence numbers, gratuitous RREPs, Hop Count, Hello Messages,

RERR and precursor lists of AODV. The source, destination and intermediate nodes of a route use data packets between source and the destination to update route lifetimes. For a bidirectional flow of data packets, no additional control packets are required and for source to destination based unidirectional flows, CONNECT packets are sent periodically by the destination towards the source to keep the forward route to the destination valid.

AODVjr has a three stage operation as follows.

1. Source node broadcasts a Route Request packet (RREQ) with a unique ID, towards the destination.

2. Destination node responds to the first Route Request with a Route Response (RREP) which is unicast back to the source along the path on which it received the Route Request, forming a unique route between the source and the destination.

3. Event is unicast from the source to the destination along the created route.

*3.1 Design of LORECOS*

*3.1.1 Protocol Extensions to AODV*

The following protocol extensions are done to AODV packet formats to implement LORECOS-AODVjr.

- RREQ (Route Request) and RREP (Route Response) packet formats are extended by adding two fields (xloc, yloc) which store the location (of the originator of RREQ) obtained during the discovery phase of AODVjr.

- A bit 'L' is added to the flag register of RREQ and RREP which indicates whether the location field (xloc, yloc) is valid.

- HELLO messages, which were removed in AODVjr are reintroduced in LORECOS for the sole purpose of localization. A bit 'H' is added to the flag register of RREP to enable the use of the RREP format for generating HELLO requests ('H' = 1) and HELLO replies ('H' = 0) instead of just HELLO messages as in AODV.

*3.1.2 Algorithm*

LORECOS-AODVjr has a three phase operation as shown in Fig. 1. It performs localization just before an event is sent to its destination. At each invocation, LORECOS tries to compute three values (DISCOVERED-LOC, CENTROID-LOC and HELLO-LOC) and advertises one of them as its location as per the algorithm given below. Before the first invocation of LORECOS, all three values are considered invalid.

**3.1.2.1 Phase 1**

1. If Event Source does not have a route to the destination,

   a) Event Source generates a RREQ and broadcasts it towards the destination with bit 'L' set to 0.

   b) Each KN which finds 'L' as 0 of the RREQ, enters its own location in (xloc, yloc) of RREQ and sets 'L' to 1 before forwarding it towards the destination.

   c) Destination copies bit 'L' and fields (xloc, yloc) from (the first) RREQ to RREP and sends RREP back to source.

   d) Source checks for bit 'L' and if set, it saves (xloc, yloc), as a valid DISCOVERED-LOC, else DISCOVERED-LOC is considered invalid.

   Else, existing value and validity status of DISCOVERED-LOC are retained.

Location Estimation with Reactive Routing in Resource Constrained Sensor Networks

### 3.1.2.2 Phase 2

1. RREQ originator broadcasts a HELLO request with 'H' set to 1 and waits for a brief (time) period T to receive HELLO replies from immediate neighbours.

2. A neighbouring node which knows its location (KN), on receiving a HELLO request, broadcasts a HELLO reply with its own location in the (xloc, yloc) field of the RREP.

3. If two or more HELLO reply packets are received, estimated location is centroid of all the received HELLOs and is called CENTROID-LOC.

   Else, if only one HELLO reply packet is received, estimated location is the location of the HELLO source and is called HELLO-LOC.

   Else, if DISCOVERED-LOC is valid, estimated location is DISCOVERED-LOC.

   Else, location estimation fails.

### 3.1.2.3 Phase 3

1. If location estimation succeeds, the location information may be sent along with the event data to the destination.

2. Since the location information with the event is a part of the payload, it is meant to be interpreted only by the destination node.

### 3.1.3 Extending AODV

In AODV, the 'D' bit of the RREQ flag register can be used to force the destination to respond to RREQs instead of intermediate nodes. Besides, the RREP flag register could also have an additional bit which indicates whether the RREP message is an AODV HELLO message or a LORECOS HELLO Request/Reply message. Thus, AODV could be readily adapted use LORECOS.

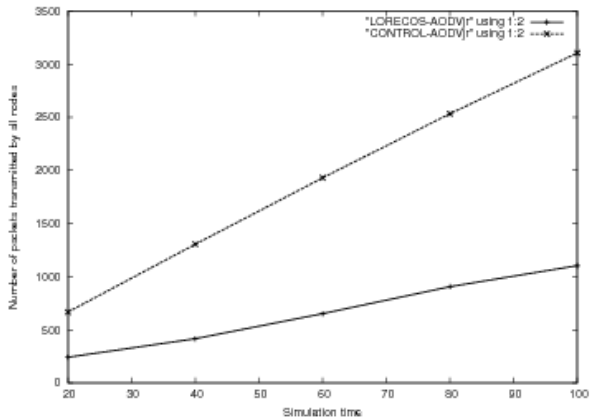

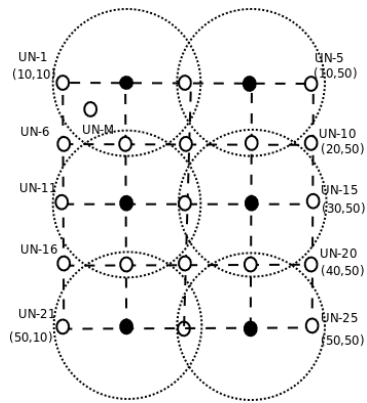

*Figure 5   Topology 2*

*Figure 4   Total number of packets in CONTROL-AODVjr Vs LORECOS-AODVjr*

*Table 2   Localization Error for UN-M*

| Algorithm | MAE(m) | RMSE(m) |
|---|---|---|
| LORECOS-AODVjr | 1.8471 | 2.2130 |
| CONTROL-AODVjr | 3.5479 | 3.9433 |

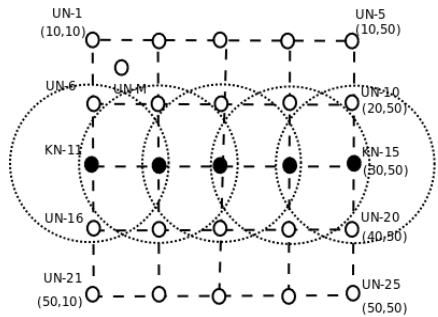

*Figure 6   Topology 3*

## 4 SIMULATION

*4.1 Design*

LORECOS-AODVjr is simulated with a custom C-based simulator. During implementation, the packet formats in [4] were maintained except for removal of unused fields from RREQ, RREP and route entries like sequence numbers, hop counts and precursor lists. Although HELLO messages are not used in AODVjr, we re-introduce them for LORECOS to be used in a special way for localization. Since we assume unidirectional flow of (sensor) events towards the destination, the destination responds with CONNECT packets towards the source as suggested in [5]. We assume a lossless channel and an ideal circular radio model. Node-id instead of IP-address, is used for addressing. We plan to implement LORECOS on a Network Simulator in future, for more fine-grained analysis.

*4.2 Topology*

For simulation we assume a field of size 40m x 40m with equally spaced 25 nodes placed in a grid as shown in Fig. 2. All nodes (whether mobile or fixed) are identical and have an ideal circular radio range of 12 m to enable routing through the grid nodes. For clarity, we analyze the performance of one mobile node (UN-M) which sends an event to the destination node 25 every 2 seconds. The destination also keeps sending CONNECT packets towards UN-M every 2 seconds.

*4.3 Mobility Model*

The mobile node UN-M starts from (15, 15) and moves across and down the field as shown in Fig. 3 at a constant speed of approximately 3 m/s (1m/ 333 millisecs). This mobility model is assumed for UN-M for all purposes, unless mentioned otherwise.

## 5 PERFORMANCE

*5.1 LORECOS-AODVjr Vs CONTROL-AODVjr*

*5.2 Design of CONTROL-AODVjr*

We simulate a control localization algorithm, henceforth called CONTROL-AODVjr which operates on the lines of the algorithm discussed in [6] to decouple localization from routing. Here, all the fixed nodes on the grid periodically broadcast HELLO (Reply) messages having their location and the UN-M computes the centroid of all location information received in accumulated HELLO messages to estimate its own location. We set HELLO broadcast period to 1 second. The location is computed just before an event is sent using AODVjr i.e. every 2 seconds.

*5.3 Comparison*

LORECOS-AODVjr and CONTROL-AODVjr are simulated for 100 seconds using Topology 1 given in Fig. 2 and the results are given in Table 2. LORECOS-AODVjr performs better since it seeks just-in-time location information from the grid instead of averaging information received at a rate independent of the speed of the UN-M. We observe from Fig. 4 that that the total number of packets transmitted in CONTROL-AODVjr increases to approximately 3 times that of LORECOS-AODVjr for 100 seconds of simulation. This is due to the wasteful periodic HELLO messages from all the grid nodes in CONTROL-AODVjr.

## 6 Analysis of LORECOS-AODVjr

Here, we illustrate how LORECOS-AODVjr adapts to various network topologies given in Figs. 2, 5 and 6 with different number of KNs to meet specific objectives. The results of a simulation run for 100 seconds are given in Table 3.

*6.1 Topology 1*

In Fig. 2, we can see that all 25 KNs participate in localization whenever needed, and simulations show that for each event generated by a UN-M, it is able to find at least two HELLO replies from neighbouring KNs for calculating the centroid. Hence it also has the least MAE and RMSE of the three configurations. This topology can be used to achieve precise coarse grained localization.

*6.2 Topology 2*

For the topology in Fig. 5, the UN-M finds just a single HELLO reply for most events, from a neighbouring KN which explains the higher percentage for HLOC. Since a single HELLO reply is available for deriving location information, the MAE and RMSE are higher than the topology discussed in 6.1. This topology is suitable for getting the approximate position of mobile node(s) with a fewer number of fixed nodes for localization.

*6.3 Topology 3*

Fig. 6, shows a set of KNs partitioning the field into two regions - top and bottom. When UN-M is in the top segment and outside the range of KNs there are no neighbouring nodes to aid localization. When the node is in the route discovery phase, the RREQ packets towards UN-25 pick up the location of one of the KNs and the RREP from the UN-25 returns this location back to UN-M. In the event sent from UN-M, this DISCOVERED-LOC is the advertised location of the UN-M. The DLOC in Table 3 indicates the percentage of location estimations derived in this manner. It may be noted that when UN-M is in the bottom region and outside the range of all KNs, LORECOS is unable to get its location most of the times because the routes to UN-25 almost never have a KN on their path. Hence this topology is good for estimating the region of a segmented field in which a UN-M lies.

*Table 3  LORECOS behaviour under different node configurations*

| TOPOLOGY | MAE(m) | RMSE(m) | CLOC | HLOC | DLOC |
|---|---|---|---|---|---|
| 1 | 1.8471 | 2.2130 | 100% | 0% | 0% |
| 2 | 6.4245 | 7.0922 | 17.5% | 77.5% | 5% |
| 3 | 7.8060 | 9.7134 | 67.5% | 12.5% | 20% |

**7 CONCLUSION**

Opportunistic localization of LORECOS helps in minimizing energy consumption which is of prime importance in sensor networks. No special assumptions are made about the sensor nodes used making it attractive for implementing LORECOS with AODV or AODVjr on actual sensor nodes. The special feature of LORECOS is that it completely adapts itself to the location information available from its surroundings which even include the nodes in the network that are not its immediate neighbours.